\documentclass[preprint,1p]{elsarticle}
\usepackage[english]{babel}
\usepackage[usenames,dvipsnames]{color}
\usepackage{amssymb}
\usepackage{amsmath}
\usepackage{enumerate}
\usepackage{graphicx}
\usepackage{bm}
\usepackage{hyperref}
\usepackage{makeidx}
\usepackage{lineno}

\begin{document}

\title{A 2-D model for friction of complex anisotropic surfaces}

\author[to]{Gianluca Costagliola}
\ead{gcostagl@unito.it}
\author[to]{Federico Bosia} 
\ead{fbosia@unito.it}
\author[tn,qm,asi]{Nicola M. Pugno\corref{cor1}}
\ead{nicola.pugno@unitn.it}

\address[to]{ Department of Physics and Nanostructured Interfaces and Surfaces Centre, 
University of Torino, Via Pietro Giuria 1, 10125, 
Torino, Italy.}
\address[tn]{Laboratory of Bio-Inspired \& Graphene Nanomechanics, Department of Civil, 
Environmental and Mechanical Engineering, University of Trento, Via Mesiano, 77, 38123 Trento, 
Italy}
\address[qm]{ School of Engineering and Materials Science, Queen Mary University of 
London, Mile End Road, London E1 4NS, UK}
\address[asi]{Ket Labs, Edoardo Amaldi Foundation, Italian Space Agency, Via del Politecnico snc, 00133 Rome, Italy}

\cortext[cor1]{Corresponding author}

\begin{abstract}

The friction force observed at macroscale is the result of interactions at various lower length scales that are difficult to 
model in a combined manner. For this reason, simplified approaches are required, depending on the specific aspect to be 
investigated. In particular, the dimensionality of the system is often reduced, especially in models designed to provide a 
qualitative description of frictional properties of elastic materials, e.g. the spring-block model. In this paper, we implement 
for the first time a two dimensional extension of the spring-block model, applying it to structured surfaces and investigating 
by means of numerical simulations the frictional behaviour of a surface in the presence of features like cavities, pillars or 
complex anisotropic structures. We show how friction can be effectively tuned by appropriate design of such surface features.

\end{abstract}


\maketitle


\section{Introduction}

The frictional behavior of macroscopic bodies arises from various types of interactions occurring 
at different length scales between contact surfaces in relative motion. While it is clear that 
their ultimate origin lies in inter atomic forces, it is difficult to scale these up to the 
macroscopic level, including other typical phenomena such as surface roughness, elasticity 
or plasticity, wear and specific surface structures \cite{persson}\cite{nos}. 
Moreover, the dependency on ``external parameters'', e.g. relative velocity of the surfaces and 
normal pressure, is neglected in approximate models such as the Amontons-Coulomb law, but violations 
have been observed \cite{exp1}\cite{exp2}.

For this reasons, simplified models are required for theoretical studies and numerical 
simulations, and friction problems can be addressed in different ways depending on the specific 
aspects under consideration. In order to improve theoretical knowledge of friction, or to 
design practical applications, it is not necessary to simulate all phenomena simultaneously 
together, and a reductionist approach can be useful to investigate individual issues. Thus, despite 
the improvement in the computational tools, still in most cases is preferable to develop simplified 
models to describe specific aspects, aiming to provide qualitative understanding of the fundamental 
physical mechanisms involved.

One of the most used approaches to deal with friction of elastic bodies consist in the 
discretization of a material in springs and masses, as done e.g. in the Frenkel-Kontorova model 
\cite{fkrev}, or the Burridge-Knopoff model \cite{burr}, the latter also known as the spring-block 
model. For simplicity, these models are often formulated in one dimension along the sliding 
direction, in various versions depending on the specific application. In recent years, interesting 
results have been obtained with these models, explaining experimental observations 
\cite{rubin}-\cite{nor}. The extension to two dimensions is the straightforward improvement to 
better describe a experimental results and to correctly reproduce phenomena in two dimensions. This 
has already been done for some systems, like the Frenkel-Kontorova model \cite{tos}\cite{faso} and 
the spring-block model applied to geology \cite{brown}-\cite{giacco}, but much work remains to be 
done for friction of complex and structured surfaces.

The interest of this study lies not only in the numerical modeling of friction in itself, but also 
has practical purposes: there are many studies relative to bio-inspired materials 
\cite{baum}-\cite{li} or biological materials \cite{gecko1}-\cite{insects} that reproduce 
non-trivial geometries that can not be reduced to one-dimensional structures. 

One of the most widely used models is the one dimensional spring-block model, which was 
originally introduced to study earthquakes \cite{equake1}-\cite{xia} and has also been used to 
investigate many aspects of dry friction of elastic materials \cite{braun}-\cite{capoz3}. In 
\cite{our} we have extensively investigated the general behavior of the model and the effects of 
local patterning (regular and hierarchical) on the macroscopic friction coefficients, and in 
\cite{our2} we have extended the study to composite surfaces, i.e. surfaces with varying material 
stiffness and roughness; finally in \cite{our3} we have introduced the multiscale extension of the 
model to study the statistical effects of surface roughness across length scales. 

In this paper, we propose a 2-D extension of the spring-block model to describe the frictional 
behavior of an elastic material sliding on a rigid substrate. Our principal aim is to compare the 
results with those obtained in the one-dimensional case and to extend our study  to 
more complex surface structures, e.g. arrangements of cavities or anisotropic structures like those 
found in biological materials. The two-dimensional spring-block model allows to consider a 
more realistic situation and captures a variety of behaviors that can be interesting for practical 
applications. In particular, we emphasize that the friction coefficients of anisotropic surface 
structures depends non-trivially on the sliding direction.

The paper is organized as follows: in section \ref{sec2}, we present the model, in section 
\ref{sec3}, we discuss the main differences with the one-dimensional case and we 
explore the role of the parameters without surface structures, highlighting the phenomenology of 
the model, in section \ref{sec4}, we present the results for standard 1-D and 2-D surface 
structures like grooves and cavities, in sections \ref{sec5a} and \ref{sec5b}, we consider more complex cases of 
anisotropic surface patterning; finally, in section \ref{sec6}, conclusions and future 
developments are discussed.

\section{Model}\label{sec2}

\begin{figure}[h]
\begin{center}
\includegraphics[scale=0.4]{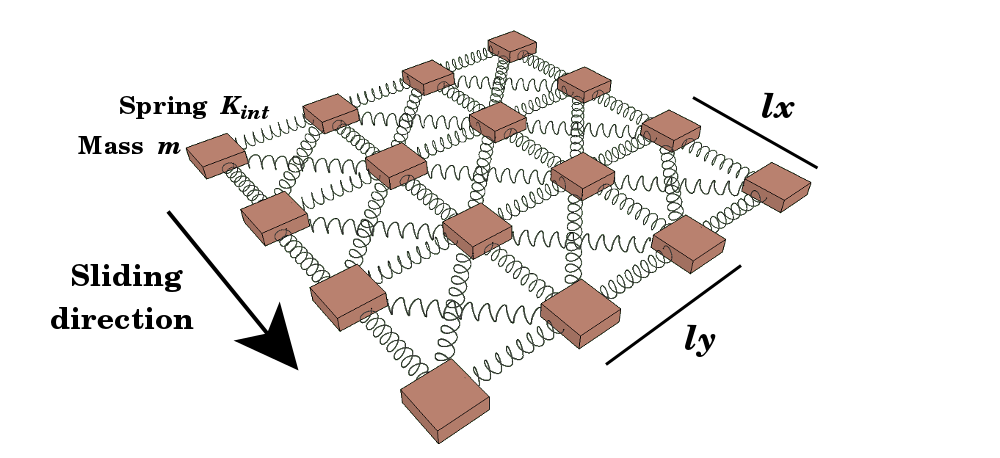}
\caption{\footnotesize Discretization of a square surface into a 2-D spring-bock model, showing the 
mesh of the internal springs. The shear springs $K_s$ attached above the blocks are not shown.}
\label{model}
\end{center}
\end{figure}

The equation of motion for an isotropic linear elastic body driven by a slider on an infinitely rigid plane with 
damping and friction can be written as: $\rho \ddot{u} = \mu \nabla^{2} u + (\lambda+\mu) \nabla (\nabla \cdot u) - 
\gamma \rho \dot{u}$, where $u$ is the displacement vector, $\rho$ is the density, $\gamma$ is the damping frequency, 
$\lambda,\mu$ are the Lam\'e constants. The 
following boundary conditions must be imposed: the top surface of the body is driven at constant velocity $v$, the bottom 
surface is subjected to a spatially variable local friction force, which we discuss below, representing the surface 
interactions between the elastic body and the rigid plane, while free boundary conditions are set on the remaining sides.

In order to simulate this system, we extend the spring-block model to the two-dimensional case: the contact surface is discretized 
into elements of mass $m$, each connected by springs to the eight first neighbors and arranged in a 
regular square mesh (figure \ref{model}) with $N_x$ contact points along the $x$-axis and $N_y$ 
contact points along in the $y$-axis. The distances on the axis between the blocks are, respectively, $l_x$ 
and $l_y$. Hence, the total number of blocks is $N_b \equiv N_x N_y$. The mesh adopted in previous 
studies of the 2-D spring-block model, e.g. \cite{olami},\cite{giacco}, does not include diagonal 
springs, but we add them to take into account the Poisson effect (our mesh in similar to that used 
in \cite{trom}).

In order to obtain the equivalence of this spring-mass system with a homogeneous elastic material 
of Young's modulus $E$, the Poisson's ratio must be fixed to $\nu=1/3$ \cite{equiv}, which 
corresponds to the plane stress case, $l_x=l_y\equiv l$ and $K_{int} = 3/8 E l_z$, where $l_z$ is 
the thickness of the 2-D layer and $K_{int}$ is the stiffness of the springs connecting the four 
nearest neighbor of each block, i.e. those aligned with the axis. The stiffness of the springs 
connecting a block with the four next-nearest neighbors, i.e. the diagonal springs, must be 
$K_{int}/2$. Hence, the internal elastic force on the block $i$ exerted by the neighbor $j$ is 
$\textbf{F}_{int}^{(ij)}  = k_{ij} (r_{ij}-l_{ij}) 
(\textbf{r}_j-\textbf{r}_i)/r_{ij}  $, where $\textbf{r}_i$, $\textbf{r}_j$ are the position 
vectors of the two blocks, $r_{ij}$ is the modulus of their distance, $l_{ij}$ is the modulus of 
the rest distance and $k_{ij}$ is the stiffness of the spring connecting them.

All the blocks are connected, through springs of stiffness $K_{s}$, to the slider 
that is moving at constant velocity $v$ in the $x$ direction, i.e. the slider vector velocity is 
$\textbf{v} = (v,0)$. Given the initial rest position $\textbf{r}_i^{0}$ of block $i$, 
the shear force is $\textbf{F}_{s}^{(i)} = K_s ( \textbf{v}t +\textbf{r}_i^{0} - \textbf{r}_i )$. 
We define the total driving force on $i$ as $\textbf{F}_{mot}^{(i)} = \sum_{j} 
\textbf{F}_{int}^{(ij)} + \textbf{F}_{s}^{(i)}$. The stiffness $K_s$ can be related to the 
macroscopic shear modulus $G = 3/8 E$, since all the shear springs are attached in parallel, so 
that by simple calculations we obtain $K_s = K_{int} l^2/l_z^2$. In the following, for simplicity 
we fix $l_z=l$. This formulation, commonly used in spring-block models, neglects 
the long-range interactions that may arise from wave propagation through the bulk 
\cite{riceruina}-\cite{hulikal}.
Here, we suppose that the local interactions are dominating, which is a reasonable
assumption for slow sliding velocities typical of the experiments we use as benchmarks \cite{baum}-\cite{li}.
This assumption has already allowed to obtain correct descriptions of the phenomena occurring
at the transition from static to dynamic friction \cite{braun}\cite{nor2}.

The interactions between the blocks and the rigid plane can be introduced in many ways: in the 
original paper  on the spring block model \cite{burr} and in earthquake related papers, e.g. 
\cite{olami}\cite{mori2}, it is introduced by means of an effective velocity-dependent force \cite{diet}\cite{ricelapus}, in 
friction studies, e.g. \cite{nor}\cite{braun}, by springs that attach and detach during motion, 
in \cite{trom}\cite{giacco} by means of the classical Amontons-Coulomb (AC) friction force. These 
various approaches give rise to slightly different quantitative results, but if they are 
implemented under reasonable assumptions, they do not significantly affect 
the overall predictions of the model, which is thought to provide a qualitative understanding of the basic mechanisms 
of friction. This is true at least for the small sliding velocities we are considering compared to the characteristic velocity
scales of the system, i.e. $l \sqrt{K_{int}/m}$. A different qualitative behavior may arise for higher sliding velocities, 
as shown for rate-and-state friction laws \cite{heaton}-\cite{elba}. In these cases, a careful evaluation
of the interplay between the friction law and sliding velocity of the system must be performed.

In this study, we adopt a spring-block model based on the AC friction force and a statistical 
distribution on the friction coefficients \cite{our}-\cite{our3}: while the block $i$ is at rest, the 
friction force $\textbf{F}_{fr}^{(i)}$ opposes to the total driving force, i.e. 
$\textbf{F}_{fr}^{(i)} = - \textbf{F}_{mot}^{(i)}$, up to a threshold value $F_{fr}^{(i)} = 
{\mu_s}_i \; F_n^{(i)}$, where ${\mu_s}_i$ is the static friction coefficient and $F_n^{(i)}$ is 
the normal force on $i$. When this limit is exceeded, a constant dynamic friction force 
opposes the motion, i.e. $\textbf{F}_{fr}^{(i)} = - {\mu_d}_i \; F_n^{(i)} 
\widehat{\dot{\textbf{r}_i}}$, where ${\mu_d}_i$ is the static friction coefficient and 
$\widehat{\dot{\textbf{r}_i}}$ is the velocity direction of the block. In the following we will 
drop the subscript $s$,$d$ every time the considerations apply to both the coefficients. 

The friction coefficients are extracted from a Gaussian statistical distribution to account for the 
randomness of the surface asperities, i.e. $p({\mu}_i) = (\sqrt{2\pi}\sigma)^{-1} 
\exp{[-({\mu}_i-(\mu)_m)^2/(2\sigma^2)]} $, where $(\mu)_m$ denotes the mean of the microscopic 
friction coefficients and $\sigma$ is its standard deviation. In order to simulate the presence of 
patterning or of structures on the surface, we set to zero the friction coefficients of the blocks 
located on zones detached from the rigid plane. The microscopic static and dynamic friction 
coefficients are fixed conventionally to $(\mu_s)_m = 1.0(1)$ and $(\mu_d)_m = 
0.50(5)$, respectively, where the numbers in brackets denote the standard deviations of their 
Gaussian distributions.

The macroscopic friction coefficients are denoted with $(\mu)_M$. The static friction coefficients 
is calculated from the first maximum of the total friction force, while the dynamic one as the 
time average over the kinetic phase. To calculate the friction coefficients as ratio between 
longitudinal force and normal force, the norm of the longitudinal force vector must be 
calculated. When calculating time averages, care must be taken in the order of the 
operations, if there is an inversion of the friction force (i.e. some blocks exceed the rest 
position, as in the analytical calculations of \cite{our}) or a periodic motion takes place, 
switching the operations of norm and time average produces different results. In these cases, 
the calculation closer to the realistic experimental procedure must be adopted. However, in the 
following results, we have checked that the above conditions do not occur and the order of the 
operations is irrelevant. The model does not include roughness variations during sliding or 
other long term effects, so that the results for dynamic friction are to be considered within the 
limits of this approximation

A damping force is added to eliminate artificial block oscillations: in \cite{braun} and in 
the papers based on it (e.g. in [39]) this is done by means of a viscous damping force 
proportional to the velocity of the block, i.e. $\textbf{F}_{d}^{(i)} = - \gamma m 
\dot{\textbf{r}_i}$. However, there is another option, e.g. in the 2-D model in \cite{trom}, where 
the damping is imposed on the block oscillations  between each pair of blocks $i$ and $j$, i.e. 
$\textbf{F}_{d}^{(ij)} = -m \gamma \; (\dot{\textbf{r}_i} - \dot{\textbf{r}_j})$, thus emulating 
the description usually adopted for viscoelastic materials \cite{hunter}-\cite{carb}. In section 
\ref{sec3}, we discuss the different behavior obtained with the two approaches, but in the 
following of the study we adopt the former one, which is the simplest to allow damping of non 
physical oscillations.

Thus, the complete equation of motion for the block $i$ is: $m \ddot{\textbf{r}_i} = \sum_j 
\textbf{F}_{int}^{(ij)} + \textbf{F}_{s}^{(i)} + \textbf{F}_{fr}^{(i)} + 
\textbf{F}_{d}^{(i)}$. The overall system of differential equations is solved using a fourth-order 
Runge-Kutta algorithm. In order to calculate the average of any observable, the simulation must be 
iterated, extracting each time new random friction coefficients. In repeated tests, an integration 
time step $h=10^{-8} s$ proves to be sufficient to reduce integration errors under the statistical 
uncertainty in the range adopted for the parameters of the system.

We consider only a square mesh, i.e. $N_x = N_y \equiv N$, and we will specify the number of blocks 
for each considered case. The default normal pressure is $P=0.05$ MPa, so that the normal force on 
each block is $F_n^{(i)} = P l^{2}$ and the total normal force is $F_n = P l^{2} N^{2}$. The slider 
velocity is $v = 0.05 $ cm/s. We will discuss in section \ref{sec3} the motivations for these 
choices, but in any case the results display small dependence on these parameters.

Realistic macroscopic elastic properties  are chosen, e.g. a Young's modulus $E = 10$ MPa, which 
it typical for a soft polymer or rubber-like material and a density $\rho = 1.2$ g/cm$^{3}$. 
The distance between blocks $l$ in the model is an arbitrary parameters representing the smallest 
surface feature that can be taken into account and it is chosen by default as $l = 10^{-3}$ cm, so 
that the order of magnitude matches those typical of surface structures used in experiments 
\cite{baum}-\cite{li}.

\clearpage

\section{Results}

\subsection{Non-patterned surface}\label{sec3}

\begin{figure}[h!]
\begin{center}
\includegraphics[scale=0.6]{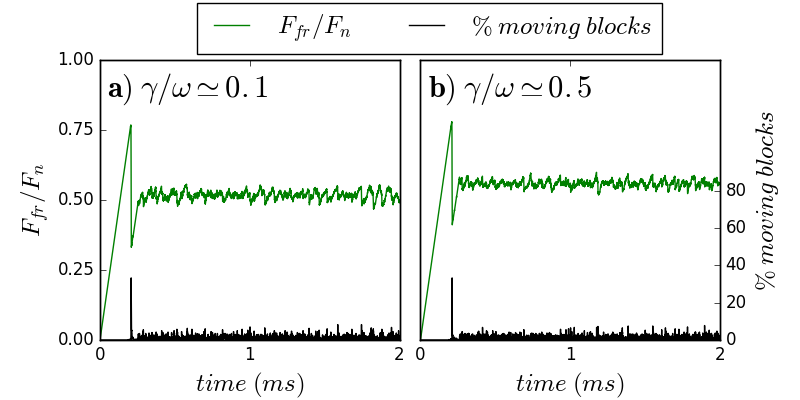}
\caption{\footnotesize Time evolution for the total friction force and percentage of moving blocks 
for $N=20$, pressure $P=0.1$ MPa, velocity $v=0.1$ cm/s, $\gamma/\omega = 0.1$ (a) or 
$\gamma/\omega = 0.5$ (b), where $\omega$ is the internal frequency $\omega \equiv 
\sqrt{K_{int}/m}$. The other parameters are set to the default values. Greater damping enhances the 
dynamic friction coefficient and reduces stick-slip oscillations.}
\label{fig1a}
\end{center}
\end{figure}

In this section, we model friction problems relative to homogeneous, non-patterned surfaces  
varying the fundamental parameters to understand the overall behavior and to compare it with that 
of the 1-D model studied in \cite{our}. In figure \ref{fig1a}, the friction force behavior as a 
function of time is shown with the default set of parameters: there is the linearly growing static 
phase, up to the macroscopic rupture event, followed by the dynamic phase in which the system 
slides with small stick-slip oscillations at constant velocity $v$. The percentage of blocks in 
motion as a function of time is also shown: in the kinetic phase, single blocks or small groups 
slip simultaneously but not in a synchronized manner with respect the rest of the surface.

\begin{figure}[h!]
\begin{center}
\includegraphics[scale=0.6]{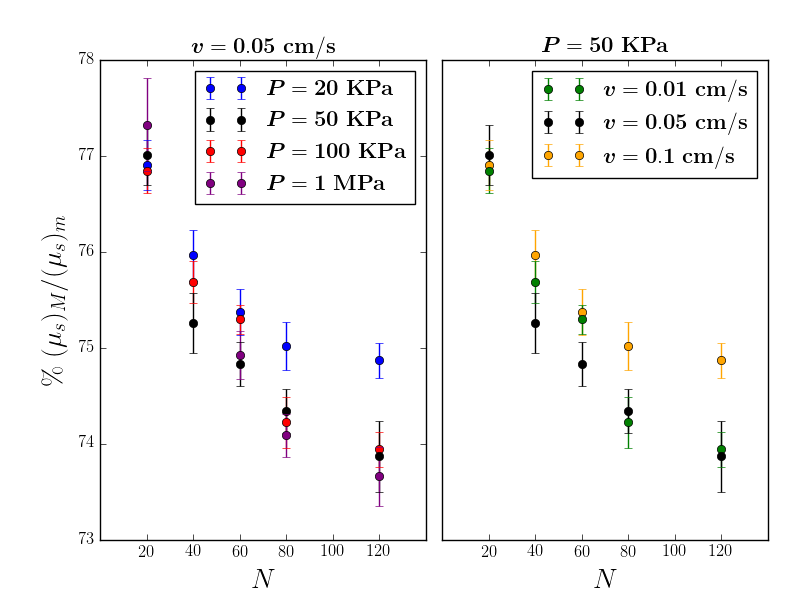}
\caption{\footnotesize Static friction coefficient as a function of the number of blocks $N$ by 
varying the applied pressure $P$ with the default velocity $v=0.05$ cm/s (a), and by varying the 
velocity $v$ with the default pressure $P=50$ KPa (b). Thus, the black dots on both sides show 
the curve for the default set of parameters. Variations with respect to this are limited to 
few percent in the typical ranges of these parameters.}
\label{fig1aa}
\end{center}
\end{figure}

The first difference with the 1-D model is that the 2D array of springs shown in figure \ref{model} 
allows to simulate the Poisson effect, i.e. a deformation in the transversal direction due to the 
stretching in the longitudinal one. Secondly, due to the model definition explained in section 
\ref{sec2}, the stiffnesses do not depend on the total number of blocks, so that increasing $N$ 
does not modify the elastic properties, but only the size of the system. Since the number of points 
grows as the square of the side, $N \gtrsim 100$ can already be considered a large system, as shown 
in figure \ref{fig1aa}, where the size effects on the global static friction coefficient are shown. 
Similar results hold for the dynamic friction. In the left panel (figure \ref{fig1aa}a) and in the 
right panel (figure \ref{fig1aa}b), the influence of the applied pressure $P$ and the slider 
velocity $v$ is also shown, respectively. In the typical ranges of these 
parameters, variations are limited within few percent, so that in the following we adopt typical 
values, e.g. $v=0.05$ cm/s and $P=50$ KPa without further discussions about their influence.

\subsubsection{Role of damping}\label{sec3a}

\begin{figure}[h]
\begin{center}
\includegraphics[scale=0.6]{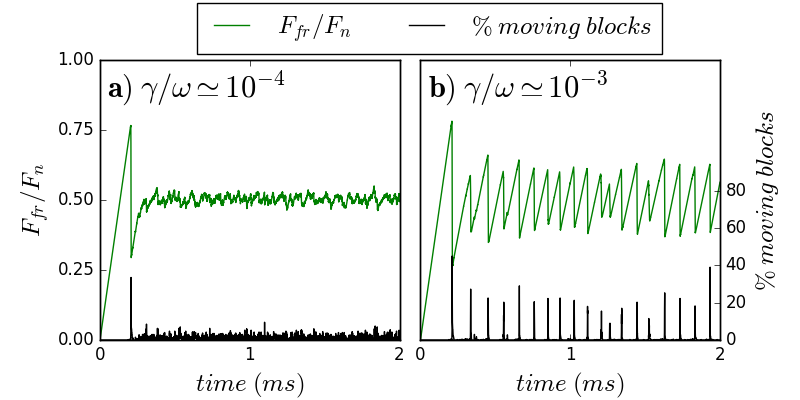}
\caption{\footnotesize Time evolution for the total friction force with the same 
parameters of figure \ref{fig1a}, except that the damping is imposed on the relative velocity 
between neighboring blocks, i.e. using a viscoelastic material model, respectively with 
$\gamma/\omega = 10^{-3}$ (a) or $\gamma/\omega = 10^{-4}$ (b), where $\omega$ is the internal 
frequency ($\omega \equiv\sqrt{K_{int}/m}$ ). The static friction coefficient remains unchanged, 
but the kinetic phase is totally different, in particular for higher damping values there are 
greater stick-slip oscillations.}
\label{fig1b}
\end{center}
\end{figure}

As mentioned, two possible approaches can be adopted to introduce viscous damping in the model. If 
we introduce a viscous damping force on the velocity, i.e. $\textbf{F}_{d}^{(i)} = - \gamma m 
\dot{\textbf{r}_i}$, there is an increase on the dynamic friction coefficient due to the damping 
which reduces the slip phases, similarly to the effect observed in \cite{our}. This does not affect 
the general behavior of the system, as long as $\gamma$ is in underdamped regime, i.e. 
$\gamma<\omega\equiv\sqrt{K_{int}/m}$.

The other option consists in assuming the damping to be dependent on the relative oscillations 
between blocks, i.e.  $\textbf{F}_{d}^{(ij)} = -m \gamma \; (\dot{\textbf{r}_i} - 
\dot{\textbf{r}_j})$, thus reproducing the generalized Maxwell model for viscoelastic materials. 
This radically changes the previously-described kinetic phase: for small damping values, there is a 
limited increase of the dynamic friction with small stick slip events, but for large 
damping, the fluctuations become larger and the kinetic phase consists in collective slips of the 
whole surface (figure \ref{fig1b}). The explanation for this is that this type of damping favours 
the elimination of relative block oscillations, enhancing the coherence of the system, so that 
sliding events can involve a large number of blocks also during the kinetic phase. 

This behavior is highly non-trivial, since it is influenced not only by the sliding velocity or the 
elastic properties of the surface, but also by the discretization parameters, i.e. the number of 
blocks $N$: for example, with $N=80$, the stick-slip oscillations are reduced, since for larger 
systems it is difficult to obtain collective slips and it is more likely that different portions of 
the surfaces move independently. 

Thus, the model can describe a variety of different situations and can capture the richness of 
behaviour of the viscoelastic material. In the following, we adopt the first solution, i.e. a 
viscous damping force on the velocity of the blocks, since it provides a simpler approach for 
damping artificial block oscillations, and we fix $\gamma = 500 $ ms$^{-1}$ ($\gamma/\omega \simeq 
0.1$).

\subsubsection{Detachment fronts}\label{sec3b}

In this section, we focus on the transition from static to dynamic friction, corresponding to 
the maximum of the total friction force and the following drop in figure \ref{fig1a}. The 
spring-block model has been used in many recent studies to obtain valuable insights on this aspect
\cite{bouch}-\cite{nor} and confirming fundamental experimental observations about the onset of the 
dynamic motion \cite{rubin},\cite{rubin2}-\cite{fineberg2}. Our aim is not a detailed study 
of the wave propagation and the rupture fronts before the sliding, for many accurate works have 
been produced on these topics \cite{svet}-\cite{unat}, but to show how the 2-D model allows to 
qualitatively predict the phenomena illustrated in the literature.

\begin{figure}[h]
\begin{center}
\includegraphics[scale=0.45]{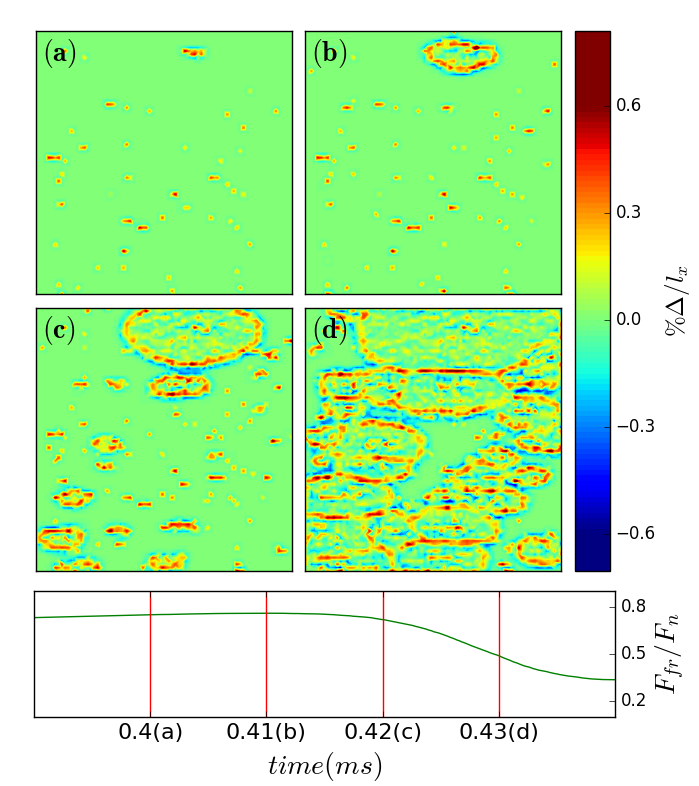}
\caption{\footnotesize Time snapshots of the spring mesh deformation $\Delta$ along the longitudinal direction 
on the surface divided by the block distance $l_x$, so that positive values (red) indicate 
compression and negative values (blue) tension. Before the maximum of the friction force is 
reached, some blocks with weak static friction thresholds detach (a), then a rupture front 
nucleates from the weakest point, corresponding to the instant of the maximum force before the drop 
(b); the front propagates while other fronts nucleate elsewhere (c) finally, the whole surface 
slides leaving a non-uniform distribution of regions under tension/compression (d).}
\label{fig2}
\end{center}
\end{figure}

In figure \ref{fig2}, four snapshots of the longitudinal deformation on the surface at different 
times of the transition are shown: starting from the points with the weakest static friction 
thresholds, rupture fronts propagate on the surface, until the whole surface slides (see 
the caption of figure \ref{fig2} for a detailed description). The maximum force, i.e. the point in 
which the global static friction coefficient is calculated, takes place when the first rupture 
front begins its propagation; then the blocks are progressively reached by the fronts and relax, 
corresponding to the phase with the drop of the friction force. This decrease ends when the whole 
surface has been reached by the rupture fronts and the overall sliding motion begins. At the 
beginning of the sliding, the spring mesh is frozen in a non-uniform 
distribution of regions of compression and tension. These regions tend to relax during the 
subsequent kinetic phase, in which different portions of the surface have continuous but incoherent 
stick-slip motion, and regions of residual stress remain. This has already been noted in the 1-D 
model \cite{braun} and observed experimentally \cite{rubin}, in terms of ''memory effects`` after 
the transition to kinetic friction \cite{nor1}. The surface deformation during the transition from static 
to dynamic friction is illustrated in Video 1 together with the time evolution of the friction force.

In 2-D models, the shape of the rupture front in the horizontal plane can be studied: before 
the nucleation of a front, the detachment propagates first to the neighbors of the weakest 
threshold point along the sliding direction, so that the the nucleation region is not a single 
point, but more likely a segment. For this reason, the fronts in figure \ref{fig2} display an 
elliptical shape. 

Many details of these simulations depend on the chosen parameters: the thresholds distribution, 
which is a way to parametrize the surface roughness, but also the velocity and the elasticity of 
the material affect the number of fronts, the speed of propagation and the duration of the friction 
force decrease. Moreover, the model does not take into account the modification of the effective 
contact area during the transition. However, it is evident that the avalanche of ruptures originate 
from the regions with weakest thresholds and then propagates to the whole surface in all the 
directions, similarly to avalanches in fracture mechanics \cite{bouch},\cite{fineberg3}. Also, it is 
interesting to note the non-trivial persistence of residual deformations in correspondence with the 
regions of interaction between multiple waves, deriving from the inelastic nature of the model.

The role of the weakest thresholds is confirmed also in \cite{our2}, where it is shown that the distribution of 
the static friction thresholds deeply affect the global static friction and the onset of motion, while it is almost irrelevant 
for the dynamic phase. Thus, in a real material the nucleation points could be the contact points with imperfect contact on the 
surface. On the basis of this observation, we discuss in the next sections how static friction can be radically modified by 
structures that give rise to non-trivial stress distributions on the surface before the sliding phase.

\clearpage

\subsection{Patterned structures}\label{sec4}

\begin{figure}[h!]
\begin{center}
\includegraphics[scale=0.35]{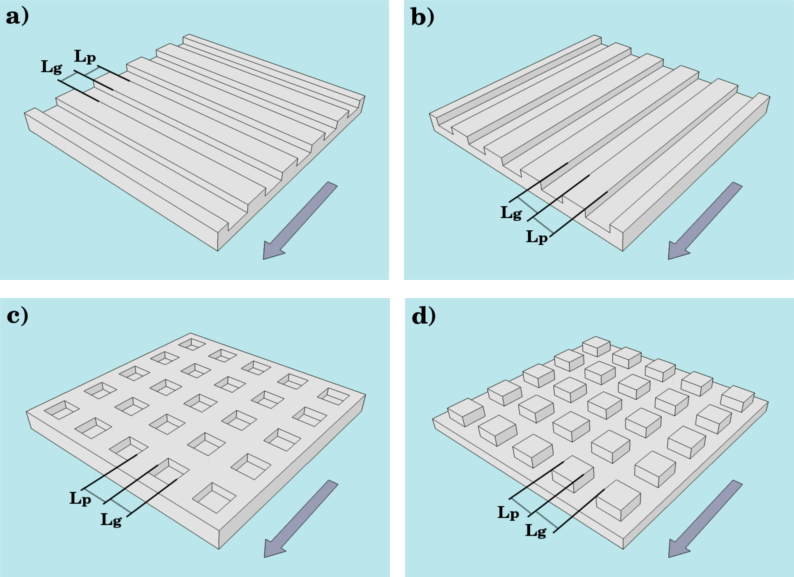}
\caption{\footnotesize Single-level surface structures considered in the simulations: patterning 
with grooves in direction perpendicular (a) or parallel (b) to the motion. Square cavities
(c) and pillars (d). The number $n_g \equiv L_g/l_x$ is the ratio between the size of the structure 
and the elementary block distance. The arrow denotes the sliding direction. The 
patterns are modelled as 2-D surfaces but graphically represented as 3-D structures for illustrative purposes.}
\label{fig3}
\end{center}
\end{figure}

First we consider single-level surface structures, i.e. described by only one characteristic length 
scale, such as those shown in figure \ref{fig3}. The 2-D surface allows to simulate 
more configurations than those studied in the one dimensional case, e.g. in \cite{our}, which is 
limited to structures similar to figure \ref{fig3}a. In experimental tests \cite{baum}, grooves 
aligned with the sliding direction, like those in figure \ref{fig3}b, have also been 
considered, while square cavities and square pillars (figure \ref{fig3}c and \ref{fig3}d, 
respectively), are the simplest two dimensional structures that we can consider. Similar structures have been 
investigated experimentally \cite{he}-\cite{grei}.

In order to simulate these structures, we set to zero the friction coefficients of the blocks corresponding to regions no longer 
in contact with the sliding plane. This is a 2-D model of the structures shown in \ref{fig3}, in which grooves correspond to 
regions without friction, while effects occurring in the depth direction are neglected, e.g. mechanical interlocking, geometric 
nonlinearities, and variability in stresses normal to the surface. However, this does not modify our general conclusions. To 
characterize the stress state of the surface, we define the surface stress field $\sigma \equiv F_{mot}/l^2$, which in the 
static phase is equivalent to the tangential stress $F_{fr}/l^2$ for the regions in contact with the substrate. In the 
following, unless otherwise stated, we indicate as ''stress`` the modulus of $\sigma$, while we denote with $\sigma_x$ and 
$\sigma_y$ its components along the x- and y-axis, respectively.

We denote with $L_g$ the width of generic non contact regions, like grooves or holes, and with $L_p$ the width of contact 
regions, like pillars or pawls, as shown in figure \ref{fig3}. The ratios $n_g \equiv L_g/l_x$ and 
$n_p \equiv L_p/l_x$ represent the number of blocks contained in these regions, which are convenient 
adimensional numbers to classify the width of the structure. In the following, if only $n_g$ is 
reported, we are considering the case $n_g = n_p$. The system parameters are fixed to the default 
values with $N_x=N_y=120$.

\subsubsection{Static Friction}

\begin{figure}[h!]
\begin{center}
\includegraphics[scale=0.5]{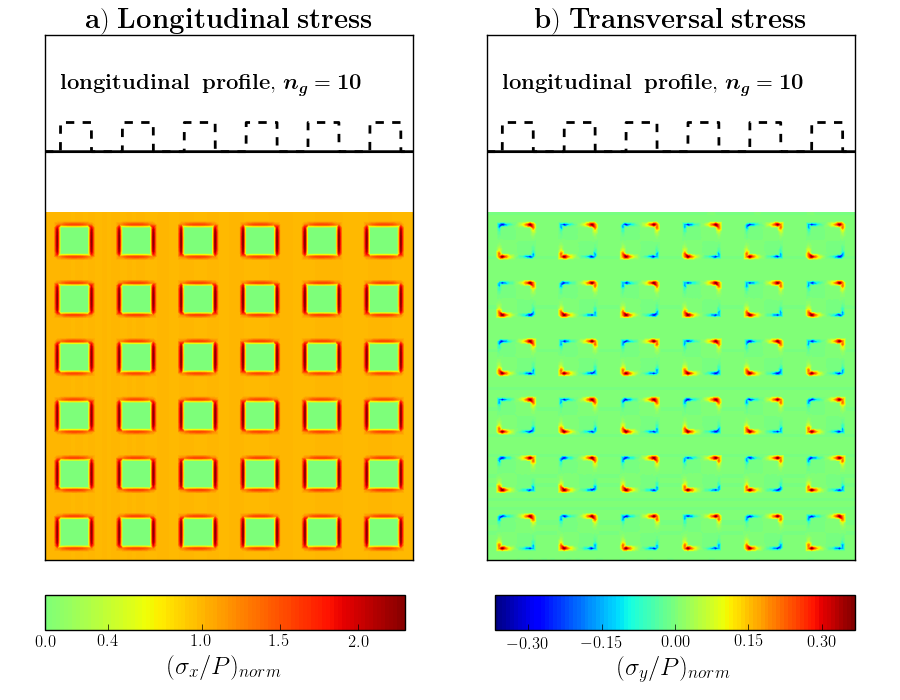}
\caption{\footnotesize Longitudinal (a) and transversal stress (b), divided by the pressure on the 
2-D surface, before the blocks motion, for a structure with square cavities as in 
figure \ref{fig3}c and $n_g=10$ (the dotted lines above shows the surface profile). The 
stress-pressure ratio is also normalized with the value obtained for a smooth surface, so that, for 
example, the normalized value is fixed to one for non-edging blocks. The stress is accumulated at 
the edge of the cavities with a non-zero component in the transversal direction. }
\label{fig4}
\end{center}
\end{figure}

In \cite{our} we have shown that in the static phase, i.e. before every block begins to slide, 
the in-plane surface stress is mostly concentrated at the edge of the grooves. Here, the same results are 
obtained and, more in general, we observe that stresses are concentrated at the edges of 
the structure in both directions, as shown in figure \ref{fig4} for the configuration of 
cavities. Due to the Poisson effect, stress components also appear in the transversal 
direction. For example, the structures in figure \ref{fig3}c tends to be deform as a trapezoid 
with the greater basis in the forward direction. Similar deformations occur in the case of 
grooves or other rectangular shapes. Vice versa, a square pillar structure such as in figure 
\ref{fig3}d deforms like a trapezoid with the smaller basis in the forward direction. Video 2 illustrates the time evolution of 
the total friction force and the longitudinal component of the surface stress distribution in the case of square cavities with 
$n_g=10$ (as in figure \ref{fig4}).

For a generic structure, the total stress is mostly concentrated where concave angles are present 
and where non-negligible stress components are present in both directions. From this we deduce that, 
other parameters being equal, a structure with a great number of concave angles and a large 
perimeter is expected to have considerably reduced static friction. Practical examples of such 
structures are presented in the next section \ref{sec5a}.  

\begin{figure}[h!]
\begin{center}
\includegraphics[scale=0.5]{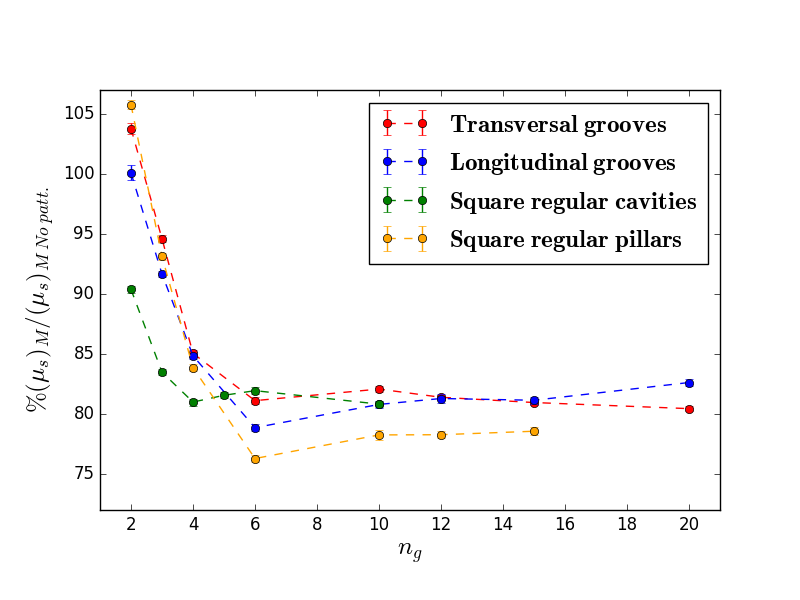}
\caption{\footnotesize Normalized static friction coefficients for the four single-level structures 
of figure \ref{fig3}. Results are normalized with respect to the static friction coefficients for a 
smooth surface (non-patterned case) and are displayed as a function of the structure characteristic 
width $n_g=n_p$. Notice the decrease of static friction for $n_g \simeq 2$ and the non 
monotonic behavior for larger sizes ($n_g > 6$). }
\label{fig5a}
\end{center}
\end{figure}

Results are shown in figure \ref{fig5a}: in the case of patterning there is the well known 
decrease in static friction for larger grooves, however in this case the behaviour is 
not monotonic. The explanation for this is that, during the rupture process, the stress is 
redistributed on the surface in a non trivial way: in the 1-D system, once the force thresholds of 
the edge blocks are exceeded, the stress is transferred only to the blocks adjacent to the edge, 
thus increasing the groove width but keeping the patterned structure. In 2-D, 
instead, ruptures can be distributed in different parts along the transversal direction, so that 
the edge formed by the attached blocks is no longer a regular patterning, but could be, for 
example, a winding profile with a non-trivial stress distribution. This influences the transition from static to dynamic 
friction and, accordingly, the maximum of the total friction force. Videos 3 and 4 illustrate the time evolution of the total 
friction force and surface stress distribution (longitudinal component) in the case of transversal and longitudinal grooves, 
respectively, with $n_g=2$.

In other terms, the crack front is forced to propagate along the pawls. When 
they are narrow, i.e. for small $n_g$, their dynamics is practically one-dimensional. If they are 
wider, the dynamics is determined by interactions of rupture fronts in different directions, so 
that the the overall behavior is more complicated and a non-monotonic dependence of static 
friction $n_g$ can arise. Moreover, before the sliding phase, 
the stress on the edges aligned with the sliding direction is slightly smaller than that on transversal ones,
but the global static friction is larger for transversal grooves with respect to longitudinal ones
(figure \ref{fig5a}). This can be only ascribed to the transition from static to dynamic friction: as noted in 
section \ref{sec3b}, the detachment front propagates first to the neighbors along the sliding direction, so that
in the case of transversal grooves, the wave propagation is hampered due to the small pawl size, despite the  
stress being slightly larger. This is less influential for large $n_g$ values and, indeed, the static friction is greater for 
longitudinal grooves. Overall, the interpretation of particular behaviors related to specific structures requires a detailed 
analysis of the onset of the dynamic motion for each specific case.

The static friction coefficient for square cavities is the smallest of the considered structures 
one for $n_g \leq 4$, but it does not decrease as significantly as for other structures again for 
larger cavities; a similar behaviour has been observed experimentally for bulk metal glass 
materials with honeycomb holes \cite{li}, suggesting that the origin of the behaviour is related to 
the stress distribution determined by its structure rather than by the material.

Finally, the square pillars with regular spacing have highest static friction for small $n_g$, but 
the smallest one for large $n_g$. The effective contact area for this structure is $S/S_{tot} = 
1/4$, so that the static friction thresholds are doubled with respect the regular patterning. 
However, for larger pillars, the stress on the edges and concave angles (contrary to hole 
structures) increases and consistently with the argument above, the friction coefficient is reduced.

The static friction of such structures is qualitatively controlled by the width of the spacings 
(in our case $n_g$) and the effective contact area as in the 1-D case, but also by its shape and
the orientation with respect to the sliding direction. In order to understand quantitatively which 
geometrical feature prevails, an accurate study of the stress distribution before the sliding and 
of the transition from static to dynamic friction is required, since in general simple 
proportionality laws cannot be formulated.

\clearpage

\subsubsection{Dynamic friction}

\begin{figure}[h!]
\begin{center}
\includegraphics[scale=0.5]{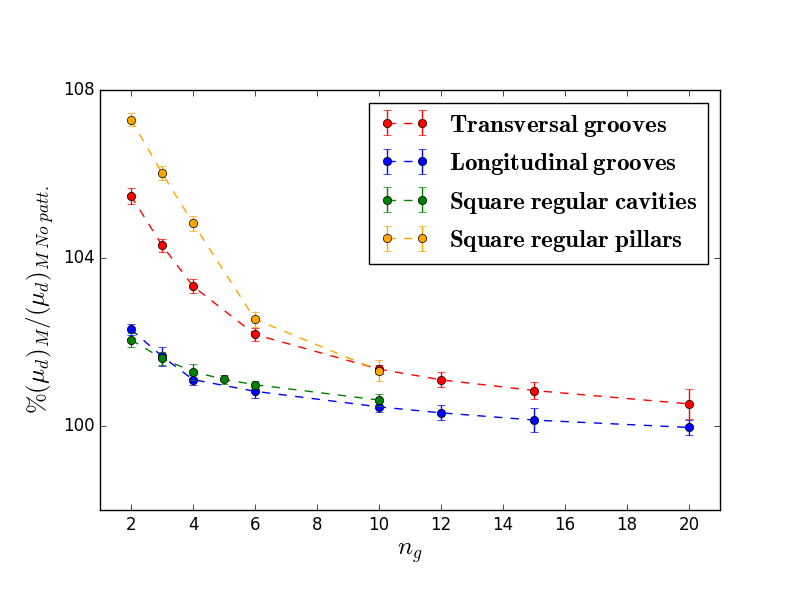}
\caption{\footnotesize Normalized dynamic friction coefficients for the four single-level 
structures of figure \ref{fig3}. Results are normalized with respect to the dynamic friction 
coefficients for a smooth surface (non-patterned case) and are displayed as a function of the 
structure typical width $n_g=n_p$. The decreasing trend with the size of the structures is limited 
to few percent with respect the non-patterned case. }
\label{fig5b}
\end{center}
\end{figure}

The dynamic friction coefficient in the presence of the considered structures displays small relative variations with respect to 
the non-patterned case. However, a trend can be observed, as reported in figure \ref{fig5b}: the dynamic friction coefficients 
are always increased with respect the non-patterned case, and are reduced by increasing the size of the structures. This can be 
explained by considering that in this regime the dynamic motion entails the non-synchronized sliding of different parts of the 
surface, with an equilibrium between moving and stationary blocks. If the level of stress increases, there are more blocks 
moving and fewer subjected to static friction, so that the sum of the friction forces during sliding, which determines the total 
dynamic friction coefficient, decreases with $n_g$.

Comparing the four different structure types, the dynamic friction coefficients increases by 
reducing the effective contact area, as noted in \cite{our}, but the geometry is also 
influential: the different behavior for longitudinal and transversal grooves, as explained for static
friction, influences also the dynamic friction due to the blocks at rest during the dynamic phase.

\clearpage

\subsection{Winding tread patterns}\label{sec5a}

\begin{figure}[h!]
\begin{center}
\includegraphics[scale=0.3]{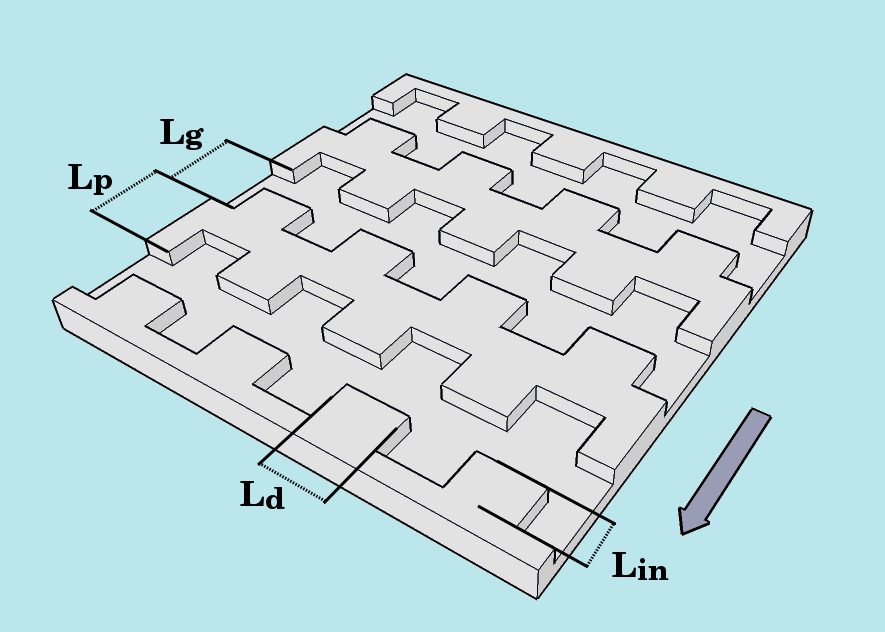}
\caption{\footnotesize Structure derived from that in figure \ref{fig3}a, in which the straight 
edge of the grooves has been modified to a winding profile with ratchets of width $L_d$ and 
depth $L_{in}$. The effective contact area is halved, like in the case of regular patterning with 
grooves and pawls of the same size.}
\label{fig6}
\end{center}
\end{figure}

As observed in the previous section, with a general non-trivial surface structure, the stress 
concentrations before the sliding is distributed at the edges and at the concave 
angles, so that for winding tread patterns we expect reduced static friction. This is confirmed 
by simulation on structures such as those shown in figure \ref{fig6}, in which the real contact 
area is the same of equal spaced grooves in figure \ref{fig3}a,b, but concave angles and 
perimeter are increased due to the winding profile of the grooves. 

As observed in \cite{our}, the effective contact area and the width of the spacings affect 
static friction too. Thus, in order to design a surface with a desired static 
friction coefficient, all of these three factors need to be considered. We consider for 
simplicity the case of figure \ref{fig6} varying the size of the features: as in the previous 
section, $L_g$ is the spacing between two consecutive structures along the sliding direction and 
$L_p$ is the width of the structure. $L_{d}$ is the width of the ratchets in the transversal 
direction to the sliding one and $L_{in}$ is their indentation. By dividing these values by $l_x$, 
the values $n_{d}$ and $n_{in}$ are obtained, corresponding to the number of blocks 
for each feature in the width and length direction, respectively. 

\begin{figure}[h!]
\begin{center}
\includegraphics[scale=0.45]{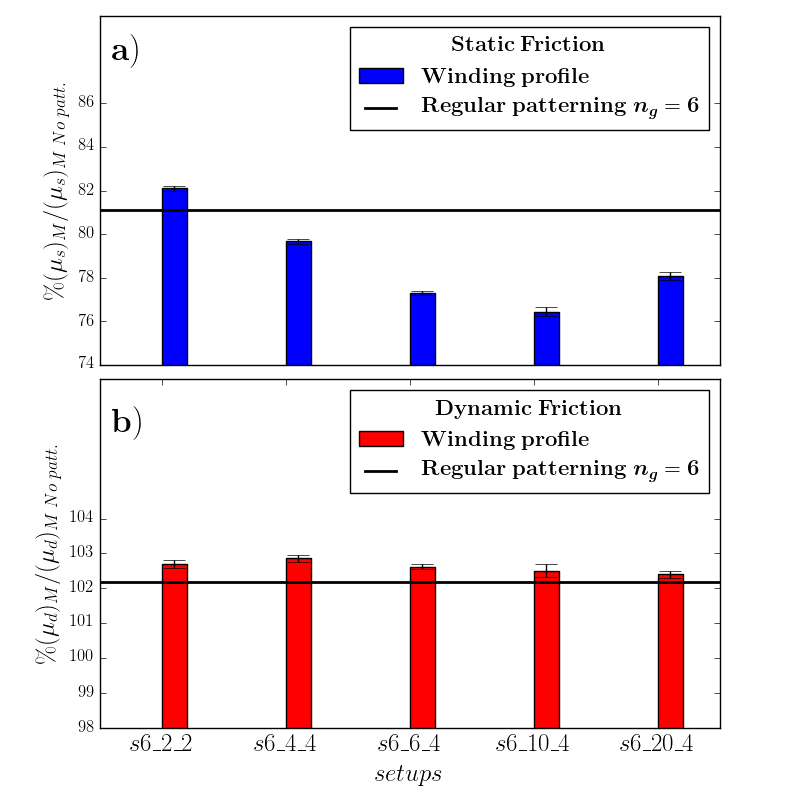}
\caption{\footnotesize Normalized static (a) and dynamic (b) friction coefficients for the 
different tread patterns in table \ref{tab7} compared to those for a regular patterning 
with $n_g=n_p=6$ (black line). The static coefficient can be further reduced with respect to the 
case of periodic regular patterning.}
\label{fig7a}
\end{center}
\end{figure}

\begin{figure}[h!]
\begin{center}
\includegraphics[scale=0.45]{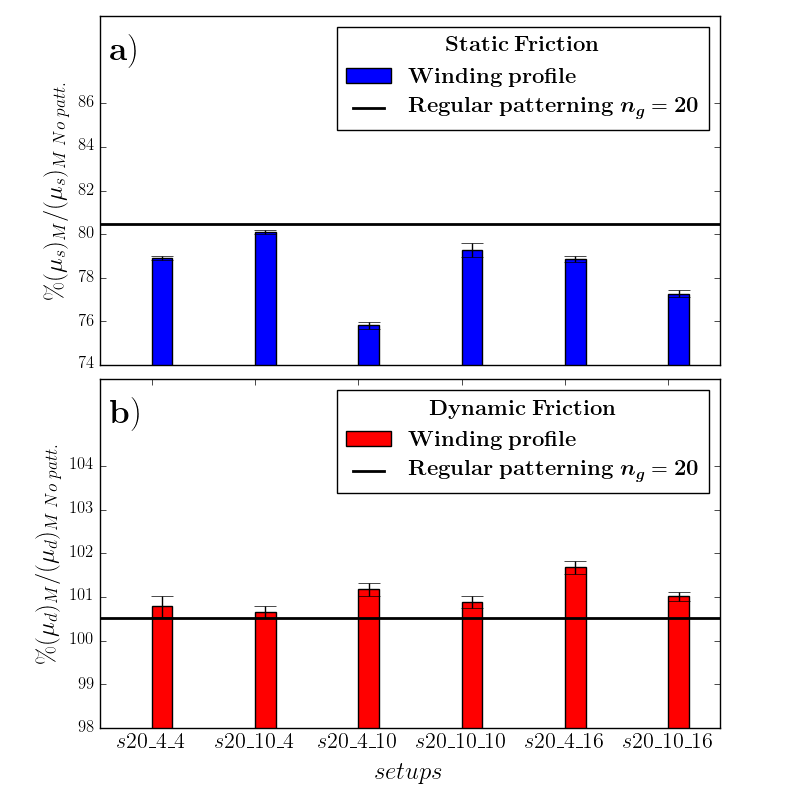}
\caption{\footnotesize Normalized static (a) and dynamic (b) friction coefficients for the 
different tread patterns in table \ref{tab7} compared to those for a regular patterning 
with $n_g=n_p=20$ (black line). While the dynamic friction coefficient displays little variation, 
the static friction coefficient can be remarkably reduced with an optimal combinations of 
parameters.}
\label{fig7b}
\end{center}
\end{figure}

In figures \ref{fig7a} and \ref{fig7b} the friction coefficients of the various tread patterns are 
shown and, in the table \ref{tab7}, their legend is reported. As expected from the previous 
discussion, static friction can be further reduced with respect to the case of periodic regular 
patterning with an increase of the perimeter and of the concave angles of the structures. Moreover, 
the precise value can be manipulated by varying the ratio between depth and width of structure, 
exploiting the high degree of tunability. There is an optimal configuration to obtain the maximum 
friction reduction, which involves ratchets whose depth is different than the width 
(e.g. configurations $s6\_10\_4$ and $s20\_4\_10$ of table \ref{tab7}). The dynamic friction can 
also be manipulated, although the relative variations are smaller. Contrarily to the static 
friction case, these structures can enhance dynamic friction with respect to the corresponding 
periodic regular structure.

Finally, by rotating the sliding direction perpendicularly to that shown in figure \ref{fig6}, 
similar qualitative considerations hold, though numerical results vary. For the configurations we 
have tested, only the $s20\_4\_10$ has the weakest static friction for both the direction. Thus, we 
can conclude that, by rotating these structures, results are not symmetric, but a configuration 
with weak the static friction coefficients in both direction can be found.

\begin{table}[h!]
\begin{center}
\begin{tabular}{|l|c|c|c||l|c|c|c|}
\hline
tread pattern & grooves $n_g$  & width $n_d$ & depth $n_{in}$ &  tread pattern & grooves $n_g$  & 
width $n_d$ & depth $n_{in}$\\
\hline
$s6\_2\_2$ & 6 & 2 & 2 &  $s20\_4\_4$ & 20 & 4 & 4 \\
\hline
$s6\_4\_4$ & 6 & 4 & 4 & $s20\_10\_4$ & 20 & 10 & 4 \\
\hline
$s6\_6\_4$ & 6 & 6 & 4 &  $s20\_4\_10$ & 20 & 4 & 10 \\
\hline
$s6\_10\_4$ & 6 & 10 & 4 &  $s20\_10\_10$ & 20 & 10 & 10 \\
\hline
$s6\_20\_4$ & 6 & 20 & 4 &  $s20\_4\_16$ & 20 & 4 & 16 \\
\hline
 &  &  &  &  $s20\_10\_16$ & 20 & 10 & 16 \\
\hline
\end{tabular}
\caption{\footnotesize Table reporting the setups of the structure of figure \ref{fig6} 
corresponding to the results presented in figures \ref{fig7a} and \ref{fig7b}. For all the setups 
only $n_g$ is reported since $n_p=n_g$.}
\label{tab7}
\end{center}
\end{table}

\subsection{Anisotropic patterns}\label{sec5b}

\begin{figure}[h!]
\begin{center}
\includegraphics[scale=0.5]{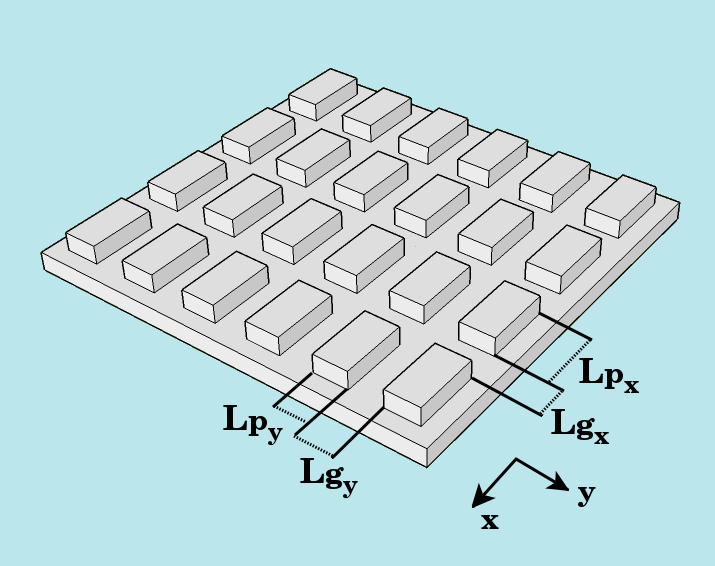}
\caption{\footnotesize Surface with rectangular pillars of size ${L_p}_x$, ${L_p}_y$ and placed at ${L_g}_x$, ${L_g}_y$
along the $x$ and $y$ direction, respectively. This simple configuration displays interesting properties due to 
the anisotropy by switching the sliding direction between the $x$ and $y$ axis. }
\label{fig8}
\end{center}
\end{figure}

In section \ref{sec4}, we discussed the different behavior obtained for longitudinal and transversal grooves, i.e. by 
rotating the grooves with respect to the sliding direction. In this section, we further investigate the role of 
anisotropic surface structures by considering, for example, rectangular pillars, as shown in figure \ref{fig8}. \\

\begin{figure}[h!]
\begin{center}
\includegraphics[scale=0.55]{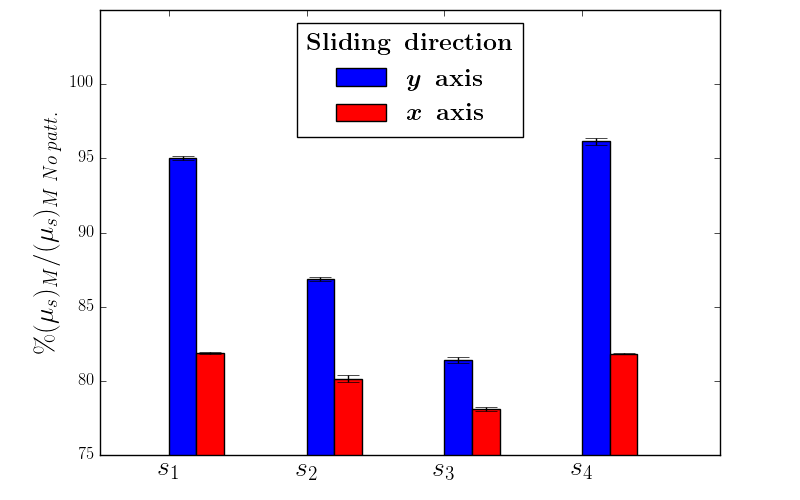}
\caption{\footnotesize Static friction coefficients for different sizes
of anisotropic pillars in \ref{tab8}, normalized by the value for a smooth surface. The $x$ and $y$ axis
are defined as in figure \ref{fig8}, i.e. the larger sides of the rectangular pillars are aligned with the $x$ axis.
There is a remarkable difference between the static friction coefficients in perpendicular sliding directions.}
\label{fig9}
\end{center}
\end{figure}

By exploiting the mechanisms observed in section \ref{sec4}, we find that with this structure static friction can vary
significantly by rotating the sliding direction. Results are reported in figure \ref{fig9}, while in table 
\ref{tab8} the size of the sides are summarized. The pillar sides are denoted with ${L_p}_x$, ${L_p}_y$ and their 
distances with ${L_g}_x$, ${L_g}_y$ along the $x$ and $y$ axis, respectively. Dividing these by the length $l$, we obtain the ratios 
${n_p}_x$, ${n_p}_y$, ${n_g}_x$, ${n_g}_y$, respectively. For rectangular pillars aligned with the sliding direction 
there is a remarkable reduction of static friction. Despite this result being intuitive, it is interesting to note the large
difference in static friction that is exclusively due to the rotation of the sliding direction. 
The anisotropy of the structure and the underlying 
mechanisms occurring at the onset of the sliding determine this behaviour. Thus, it appears that, 
to manipulate static friction with the sliding direction, anisotropic dimensions of the structure are more 
effective than complex shapes.

Additionally, we observe that, by increasing the size of the pillars, static friction decreases (as expected), and
that the differences between the two directions are also reduced. This confirms that the effect is due to the
mechanisms occurring during the transition from static to dynamic friction, as explained for grooves \ref{sec4}.\\

\begin{table}[h!]
\begin{center}
\begin{tabular}{|l|c|c|c|c|}
\hline
data set & ${n_p}_x$ & ${n_p}_y$ & ${n_g}_x$ & ${n_g}_y$  \\
\hline
$s_1$ & 8 & 2 & 4 & 4 \\
\hline
$s_2$ & 12 & 3 & 6 & 6  \\
\hline
$s_3$ & 16 & 4 & 8 & 8  \\
\hline
$s_4$ & 8 & 2 & 4 & 6 \\
\hline
\end{tabular}
\caption{\footnotesize Characteristics of the anisotropic pillars of figure \ref{fig8} 
corresponding to the results presented in figure \ref{fig9}. We denote with ${n_p}_x$, ${n_p}_y$ 
the sides the pillars, and with ${n_g}_x$, ${n_g}_y$ their distances along the $x$ and $y$ axis, respectively, 
expressed in number of elementary blocks.}
\label{tab8}
\end{center}
\end{table}

\clearpage

\section{Conclusions}\label{sec6}

In this paper, we have introduced a 2-D version of the spring-block model to investigate the friction coefficients of complex 
surfaces that cannot be reduced to one dimension. This model is fundamental for practical applications and to explain recent 
research results on friction of patterned surfaces in biological and bio-inspired materials. We have described the model in 
detail and presented benchmark results with a non-patterned surface, illustrating the effects of the model parameters, the 
general behavior of the system and to its consistency with results from the literature. We have also shown that interesting 
insights on friction can be obtained by investigating the transition from static to dynamic friction and the propagation of 
avalanche ruptures on the surface.

Next, we have considered simple patterned surfaces, e.g. longitudinal and transversal grooves, regular square cavities and 
pillars. Due to the Poisson effect, the in-plane surface stresses are non-zero in the transversal direction, so that structures 
like cavities deform and stretch in the forward sliding direction, while they undergo compression in the backward one, and vice 
versa for protruding structures like pillars. The surface stress is mostly concentrated at the edges and at concave angles. We 
have investigated how the friction coefficient is modified by varying the size of these structures, finding non-trivial 
behaviors that depend on the surface redistribution of stress during the transition from static to dynamic friction. The most 
interesting predictions are relative to the non-monotonic behavior of static friction by varying the size of the cavities (in 
agreement with experimental results) and the maximum static friction reduction obtained for structures with large regular square 
pillars.

Finally, we have considered winding tread patterns, which have the same contact area and the same spacings of regular groove 
patterns, but a greater number of concave angles and perimeter. As expected from the previous observations, we find a remarkable 
static friction reduction for some of these configurations. Thus, to manipulate the global static friction with structured 
surfaces, while in the 1-D case both the contact area and the width of the structures play a role, in the 2-D case the geometry 
of the edges also becomes fundamental. Fine tuning of static friction can also be achieved by varying the size of the specimens. 
Moreover, in the case of anisotropic structures like rectangular pillars, the friction coefficients can vary significantly with 
the sliding direction, which becomes an additional parameter to take into account. These kinds of predictions require a 2-D 
model such as the one presented herein that is able of capturing the non-trivial behavior of complex structures similar to those 
commonly observed in nature or employed in technological fields such as tire tread design.

\section*{Acknowledgments}
N.M.P. is supported by the European Commission H2020 under the Graphene Flagship Core 1 No. 696656 (WP14 ``Polymer 
Nanocomposites'') and FET Proactive ``Neurofibres'' grant No. 732344. G.C. and F.B. are supported 
by H2020 FET Proactive ``Neurofibres'' grant No. 732344. Computational resources were provided by the Centro di Competenza sul 
Calcolo Scientifico (C3S) of the University of Torino (c3s.unito.it) and by hpc@polito (http://www.hpc.polito.it).

\end{document}